\documentclass
[preprint,twocolumn,prl,10pt,nofootinbib,bibnotes,tightenlines]{revtex4}%
\usepackage{amsfonts}
\usepackage{amsmath}
\usepackage{amssymb}
\usepackage{graphicx}%
\newtheorem{theorem}{Theorem}

\newtheorem{corollary}[theorem]{Corollary}

\newenvironment{proof}[1][Proof]{\noindent\textbf{#1.} }{\ \rule{0.5em}{0.5em}}
\begin{document}
\preprint{UATP/08-01}
\title{Poincar\'{e} Recurrence, Zermelo's Second Law Paradox, and Probabilistic
Origin in Statistical Mechanics }
\author{P. D. Gujrati}
\email{pdg@arjun.physics.uakron.edu}
\affiliation{The Department of Physics, The University of Akron, Akron, OH 44325}
\date{\today}

\begin{abstract}
We show that Poincar\'{e} recurrence does not mean that the entropy will
eventually decrease, contrary to the claim of Zermelo, and that the
probabilitistic origin in statistical physics must lie in the external noise,
and not the preparation of the system.

\end{abstract}
\maketitle

As first pointed out by Kr\"{o}ning \cite{Kroning}, and later developed by
Boltzmann \cite{Boltzmann0}, any deep understanding of the second law of
thermodynamics in terms of entropy due to Clausius \cite{Clausius} must
involve a probabilistic approach; see \cite{ter Haar,Lebowitz} for two of the
excellent reviews. This was the first approach in physics to establish that
fundamental laws of Nature need not be strictly deterministic. However, as
many phenomena at the microscopic level such as nuclear decay also require a
probabilistic approach for their understanding, the probabilistic
interpretation is not just a consequence of a macroscopic system; yet it has
to be exploited in statistical physics. To appreciate this probabilistic
approach in statistical physics, we note that in the Gibbs formulation, the
entropy is given by the \emph{average} of the degree of uncertainty
$u_{i}(t)=-\ln p_{i}(t)$ of the $i$th\ microstate:
\begin{equation}
S(t)=\sum p_{i}(t)u_{i}(t)\geq0,\ \ \sum p_{i}(t)=1; \label{Eq_Entropy}%
\end{equation}
here $p_{i}(t)$ is the probability for the $i$th microstate at time $t,$ and
the sum is over all distinct microstates $W$. We allow the possibility that
some of the probabilities may be zero. Thus, one envisions the system to be in
different microstates with certain probabilities. To introduce the concept of
$p_{i}(t)$, we construct a Gibbs ensemble as containing $\mathcal{N}$ replicas
of the system, $\mathcal{N}_{i}$ of which are in microstate $i$. Then,
$p_{i}(t)\equiv\mathcal{N}_{i}/\mathcal{N}$ in the limit $\mathcal{N}$
$\rightarrow\infty;$ the limit will always be implicit. If there is only one
microstate $i=0$ possible \cite{Note} so that $\mathcal{N}_{0}=\mathcal{N}$,
the entropy is identically zero as the system is in $i=0$ with
\emph{certainty}. According to the second law, the entropy of an isolated
system cannot decrease. The equilibrium is attained when the entropy becomes
maximum, which occurs when all microstates have the same probability:
\begin{equation}
p_{i}(t)\rightarrow1/W. \label{Eq_Probability}%
\end{equation}
Once the equilibrium is achieved, the entropy cannot decrease if the system is
left undisturbed.

However, the application of the Poincar\'{e} recurrence theorem
\cite{Poincare}, see below, gave rise to Zermelo's \cite{Zermelo} paradox,
which has not been resolved to everyone's satisfaction yet, and is the subject
matter of this work. The recurrence theorem is valid for a classical system
and basically states that provided an isolated mechanical system, in which the
forces do not depend on the velocities of the particles, remains in a
\emph{finite} part of the phase space during its evolution, then the
uniqueness of classical trajectories implies that a given initial state must
come \emph{arbitrary close} to itself infinitely many times. Zermelo
\cite{Zermelo} argued that since the entropy is determined by the phase point,
then it must also return to its original value according to the recurrence
theorem. Thus, if the entropy increases during a part of \ the time, it must
decrease during another and this increase and decrease in the entropy must
occur infinitely many times, thereby violating the second law. In addition, it
is not just the microstate itself, but its probability of occurrence that
determines the entropy. As we demonstrate here, not appreciating this fact has
given rise to the paradox due to Zermelo. We will also show that the origin of
this probabilistic behavior in not in the method of preparation of the system,
which leaves the system deterministic; rather it lies in the stochastic
interaction with the environment, no matter how weak, for the second law to work.

According to Boltzmann \cite{Boltzmann}, recurrences are not inconsistent with
the statistical viewpoint: they are merely statistical fluctuations, which are
almost certain to occur. Indeed, Boltzmann \cite{Boltzmann}, Smoluchowski
\cite{Smoluchowski}, and others recognized that the period of a Poincar\'{e}
cycle is so much larger for a macroscopic system to be almost infinitely large
so that the \emph{violation} of the second law (decrease in entropy of an
isolated system) would be almost impossible to occur in our life. The period
of the cycle will be many orders of magnitude larger than the present age of
the universe \cite{Huang}. While an appealing argument, it is hard to
understand its relevance as the argument compares presumably a
system-intrinsic time, the recurrence time, with a system-extrinsic time, the
time of observation or the age of the universe. Moreover, the recurrence
theorem is valid for a \emph{deterministic system} as will be detailed below,
while the second law is valid for a stochastic system requiring a
probabilistic approach which necessiates exploiting an ensemble . In
particular, Poincar\'{e}'s recurrence theorem states that the mechanical
system will revisit the neighborhood of its initial state with
\emph{certainty} (with probability $p=1$), while for a statistical system, the
probability of revisit is extremely small (indeed $p\simeq1/W$ for a
macroscopically large system) as we discuss below.%

\begin{figure}
[ptb]
\begin{center}
\includegraphics[
trim=0.668835in -0.057331in -0.048335in 0.000000in,
height=4.2082in,
width=2.0141in
]%
{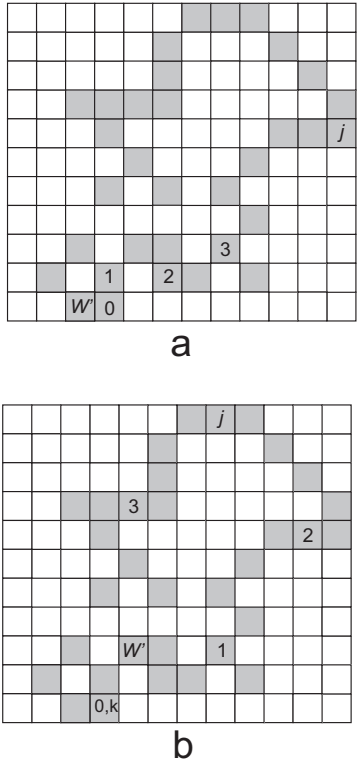}%
\caption{Schematic evolution of an initial microstate $0$ as a function of
time in the phase space. A cell shows a microstate and the numbers show the
sequential emrgence of microstates in time. (a) shows not only the unique
deterministic evolution, but also one of many possible stochastic evolution.
Being a deterministic evolution, no microstate recurs except the initial one
at time $t_{\text{R}}.$ (b) shows one of the many possible stochastic
evolution in which the initial microstate recurs at time $t_{\text{k}%
}<t_{\text{R}}.$}%
\label{fig_microstates_poincare_1.ps}%
\end{center}
\end{figure}

We first discuss the recurrence theorem for completeness and then its
relevance for the second law.

\begin{theorem}
\label{Th_Poincare_Recurrence}\textbf{Poincar\'{e} Recurrence }A microstate of
a finite classical system evolving deterministically and confined to a finite
region of the phase space during its evolution recurs infinitely many times.
\end{theorem}

\begin{proof}
We consider a classical system consisting of $N$ particles, which we take to
be point-like for simplicity, with energy $E$ in a volume $V$. We restrict
$N,E$ and $V$ to be finite to ensure that the system moves in a \emph{finite}
region of size $\left\vert \Gamma_{0}\right\vert $ in the phase space; see the
shaded region in Fig. \ref{fig_microstates_poincare_1.ps}. A microstate is
commonly defined not by a point in the phase space, but by a small volume
(shaded cells in Fig. \ref{fig_microstates_poincare_1.ps}) of the size
$\tau_{0}\equiv h^{2r}$; $h$ is Planck's constant and $r=3N$ \cite{Note0}. The
number of distinct microstates $W$ is%
\[
W\equiv\left\vert \Gamma_{0}\right\vert /\tau_{0},
\]
which is exponentially large of the order of $c^{N},$ with $c\geq1$ some
constant$.$ We assume, for simplicity, that the time required for a microstate
to evolve into a different microstate is some constant $\Delta$, and observe
the system at times $t=t_{j}=j\Delta,j=1,2.\cdots,$ to determine the
microstates. The dynamics of an isolated system with a given Hamiltonian is
completely deterministic \cite{Note01}: an initial microstate $i$ evolves in a
\emph{unique} fashion into a microstate $i_{j}$ at time $t=t_{j}$, which we
represent by the \emph{one-to-one} mapping $i\rightarrow i_{j}$. Due to the
unique evolution, the system visits each of the $W$\ microstates in time
without \emph{repeating} until it has visited all of them. It will then
revisit the initial microstate (it cannot visit any other microstate because
of the unique evolution) and then repeat the entire sequence $\left\{
i_{j}\right\}  $ exactly in the same order over and over. We show in Fig.
\ref{fig_microstates_poincare_1.ps}a the deterministic evolution of the
initial microstate $i=0$ through microstates $0_{j}$ at $t=t_{j},$ shown
schematically by $j=$ $1,2,\cdots,W^{\prime}\equiv W-1$. The next microstate
at $j=W$ will be $0$ \cite{Note1}, and the entire ordered sequence $\left\{
0_{j}\right\}  $ will be visited during the next cycle. The recurrence time
$t_{\text{R}},$ also known as the Poincar\'{e} cycle, is given by
$t_{\text{R}}\equiv W\Delta.$ Each microstate will be revisited several times
in a time $t\gg t_{\text{R}}$. This proves the recurrence theorem \cite{Note1}.
\end{proof}

A more general proof can be found in \cite{Huang}.

\begin{theorem}
\label{Th_Poincare_Cycle_Entropy}The entropy in a Poincar\'{e} cycle remains
constant so that the second law is never violated.
\end{theorem}

\begin{proof}
Consider Fig. \ref{fig_microstates_poincare_1.ps}a. Since the system is with
certainty in only one microstate $j(\operatorname{mod}W)$ at instant $t=t_{j}%
$, its entropy $S(t_{j})=0$ identically for all $j$. As $S(t_{j})$ can never
decrease, the phenomenon of recurrence does not violate the second law.

The same conclusion is also obtained in the ensemble approach. We prepare each
replica in the same microstate $0$ initially and follow its evolution in time.
Because the evolution is deterministic, each replica in the ensemble will be
in the same microstate $0_{j}$ at $t=t_{j}.$ Thus, $p_{i}(t_{j})=\delta_{ij},$
which again gives $S(t_{j})=0$. Let us now consider the system to be initially
in a "macrostate" consisting of two possible microstates $0$ and $1$ with
probabilities $p_{0}$ and $p_{1}\equiv1-$ $p_{0},$ respectively. This can also
be done for a quantum system. There are $\mathcal{N}p_{0}$ replicas in the
microstate $0$, and $\mathcal{N}p_{1}$ in the microstate $1$, and the initial
entropy is $S(0)=-$ $p_{0}\ln$ $p_{0}-$ $p_{1}\ln$ $p_{1}.$ Since the
evolution $(0\rightarrow0_{j}$, $1\rightarrow1_{j})$ at some later time
$t_{j}$ is deterministic, all the $\mathcal{N}p_{0}$ replicas are in
microstate $0_{j}$, and the remaining $\mathcal{N}p_{1}$ in microstate $1_{j}%
$, so that $\Pr(0_{j})=p_{0},$ and $\Pr(1_{j})=p_{1}.$ Consequently,
$S(t_{j})=S(0)$ so the entropy remains constant. It is easy to extend the
calculation to an initial "macrostate" consisting of any number of microstates
$i$, in particular all microstates $W$, with probabilities $p_{i}$ with the
same conclusion that the entropy given by (\ref{Eq_Entropy}) remains constant
during the Poincar\'{e} cycle. This completes the proof.
\end{proof}

The above conclusion is consistent with \emph{time-reversal invariance} in a
deterministic dynamics. As the evolution $i\rightarrow i_{j}$\ is one-to-one,
it can be inverted at any time. Thus, the forward evolution $i\rightarrow
i_{1}\rightarrow i_{2}\rightarrow\cdots\rightarrow i_{j}$ of an initial
microstate can be uniquely inverted to give $i_{j}\rightarrow i_{j-1}%
\rightarrow i_{j-2}\rightarrow\cdots\rightarrow i_{1}\rightarrow i$, and we
recover the initial microstate in this reversal. The entropy in this reversal
remains constant to ensure time-reversal invariance.

It should be commented that the non-zero initial entropy for a deterministic
system (classical or quantum) considered above is due to our mode of
preparation. It is due to our ignorance about the system and does not
represent an intrinsic property of the system. The notion of probability here
is brought into the discussion due to the preparation of the system, and we
have total control to change its probabilistic nature and to change the
entropy so that \emph{the latter is not an intrinsic characteristic of the
system}. This entropy of the deterministic system can be readily changed to
zero by making a precise measurement to determine which microstate the system
is in \cite{Note}. This "collapse" of the "macrostate" means that once the
system is known to be in a particular microstate with certainty after
measurement, its entropy will remain zero as shown above for ever, even though
the system is not in equilibrium. This is not how we expect the thermodynamic
entropy to behave. Moreover, being a constant, this entropy will never become
the maximum possible equilibrium entropy $\ln W$, unless it is already at the
maximum. This gives us the following

\begin{corollary}
A deterministic system will never equilibrate if it was not in equilibrium initially.
\end{corollary}

Ideal gases confined by idealized walls to form isolated systems have no
mechanism to achieve equilibrium, and will remain in non-equilibrium states
for ever if they were so initially; see footnote 6 in \cite{ter Haar}. For the
concept of entropy to be useful requires it to be an intrinsic property of the
system which should not be affected by the measurements if we wait long enough
after the measurements. Thus, the concept of entropy requires a particular
kind of probabilistic approach in which the evolution must not be
deterministic; rather, it must be $\emph{stochastic}$. Even an isolated system
is not truly deterministic in Nature. A real system must be confined by a real
container, which cannot be a perfect insulator. Even the container will
introduce environmental noise in the system. Thus, there are always stochastic
processes going on in a real system, which cannot be eliminated, though they
can be minimized. In the case the external noise is too strong, then there is
no sense in not considering the environment as part of the system for its
thermodynamic investigation. It is the limit in which the external noise is
too weak that is relevant for a sensible thermodynamical description of a
system, so that the external noise will not alter the average properties such
as the average energy of the system \cite{Pdg2008b}. For quantum systems, this
requires considering the Landau-von Neumann density matrix, rather than
eigenstates; see, for example, \cite{Huang}. The derivation in \cite{Huang}
clearly shows the uncertainty introduced by the presence of the environment.
The latter is not part of the system, just as in thermodynamics. It is only in
this case that the entropy will vary as the probabilities of various
microstates change in time, as we describe below.

The stochasticity introduces a new time scale $\Delta^{\prime}$ over which the
system evolves deterministically as above. Over this time-period, the the
mapping $i\rightarrow i_{j}$ is one-to-one and can be inverted to study
time-reversal. The entropy remains constant during this period. At the end of
each time period $\Delta^{\prime},$ i.e. at time $t=t_{k}^{\prime}\equiv k$
$\Delta^{\prime},k=1,2,\cdots$, the current microstate $i$ will undergo a
stochastic "jump" ( shown by the double arrow $i\twoheadrightarrow j$) to any
of the $W$ microstates $j$ \cite{Note3} brought about by the environmental
noise. We take these "jumps" to occur instantaneously just for simplicity. The
"jump" may create a new microstate not generated so far, or bring it back to a
previously visited microstate, including the initial microstate, generated
during its deterministic evolution. Such a jump to a previously generated
microstate (not the initial microstate) would have been forbidden in a
deterministic evolution alone as noted above. Many such stochastic "jumps" are
needed to bring the system to equilibrium, which requires a time interval
$t_{\text{eq}}$, so that $\Delta^{\prime}<t_{\text{eq}}$. The presence of
stochastic "jumps" give rise to a probabilistic nature to the microstates,
their probability of occurrence changing with time. This in turn changes the
entropy with time whenever "jumps" occur.

How $\Delta^{\prime}$ relates to the timescale $\Delta$ depends on the
strength of the noise; here we will assume $\Delta^{\prime}\gtrsim\Delta$,
which can be reversed without affecting our conclusions. For an isolated
system, $\Delta^{\prime}\rightarrow\infty,$ which is consistent with our
Corollary that the deterministic evolution cannot bring about equilibration
(entropy maximization). The external noise causes the entropy to increase with
time if the initial state was out of equilibrium as shown elsewhere
\cite{Pdg2008b}. Therefore, we now turn to the stochastic evolution to make
contact with the second law.

We do not have to consider the actual nature of the noise; all that is
required is its presence. The actual nature will only determine the value of
$\Delta^{\prime}$, but not the final equilibrium state, which remains
oblivious to the actual noise. This is what allows the statistical mechanical
approach to make predictions about the equilibrium state. We consider an
ensemble of $\mathcal{N}$\ replicas, each replica being identically prepared
in the same microstate $0$, so that $p_{i}(0)=\delta_{i0}$. Consequently,
$S(0)=0$. This obviously represents an extreme non-equilibrium situation.
However, since the evolution is stochastic, a microstate $i$ makes a "jump" to
another microstate $j$ ($i\twoheadrightarrow j$) or remain the same
($i\twoheadrightarrow i$) caused by the noise. It is also possible to have
$i\twoheadrightarrow j\twoheadrightarrow i$, as shown in\ Fig.
\ref{fig_microstates_poincare_1.ps}b, where we show that the system leaves the
original microstate $0$ but comes back to it at $t_{k}$. Thus, the recurrence
can happen at any time $t\geq\Delta^{\prime}$, albeit without certainty
(probability $p_{0}(t)<1$) and has no particular significance or relevance for
the Poincar\'{e} cycle for a finite system, where recurrence occurs with
certainty. This distinction in the probability of recurrence is very
important, as the entropy is determined by the probability. Just because the
initial microstate has recurred does not necessarily mean that the entropy has
reversed to its initial value $S(0)=0$, contrary to the claim by Zermelo. One
needs to consider its probability also. To establish this, we proceed as follows.

At $t_{1}^{\prime},$ there will be $\mathcal{N}p_{i}(t_{1}^{\prime}%
)$\ replicas in the $i$th microstates. In particular, there is a non-zero
probability $p_{0}(t_{1}^{\prime})<1$ that the system will be back in its
original microstate $0$. However, this in no way means that the average degree
of uncertainty $S(t_{1}^{\prime})$ has reduced to zero, as it is obtained by
(\ref{Eq_Entropy}), which requires a sum over \emph{all} microstates that are
present in the ensemble. The degree of uncertainty of the initial microstate%
\[
u_{0}(t_{1}^{\prime})=-\ln p_{0}(t_{1}^{\prime})>0,
\]
so that $S(t_{1}^{\prime})>p_{0}(t_{1}^{\prime})u_{0}(t_{1}^{\prime})>0$. The
entropy has increased. For $t_{1}^{\prime}<t<t_{2}^{\prime},$ each replica
evolves deterministically so that the entropy remains constant, as follows
from Theorem \ref{Th_Poincare_Cycle_Entropy}. This is true during each of the
intervals $t_{k}^{\prime}<t<t_{k+1}^{\prime},$ with the entropy changing at
$t_{k}^{\prime}$ as the probabilities $p_{i}(t_{k}^{\prime})$ change. During
all this time, the system has a non-zero probability $p_{0}(t)<1$ to be in the
initial microstate $0$. Eventually, the system equilibrates when
(\ref{Eq_Probability}) holds so that
\[
u_{i}(t)\rightarrow\ln W,\ \ \ i=0,1,\cdots,W^{\prime},
\]
which is exactly the entropy $S(t)=\ln W$ of the system obtained by summing
over all microstates; see (\ref{Eq_Entropy}). We observe that \emph{in
equilibrium, the entropy is exactly the degree of uncertainty of any
microstate and, in particular, the initial state}.

Thus, we come to the following theorem:

\begin{theorem}
Even for a stochastic evolution, which is needed for a statistical system, the
recurrence of the initial microstate does not violate the second law.\ 
\end{theorem}

The entropy remains constant after equilibrium is reached. The recurrence of
the initial microstate $0$ (but with $p_{0}(t)=1/W$ ) in the stochastic case
does not mean that the entropy reverts to the initial entropy $S(0).\ $On the
other hand, the \emph{true recurrence} of the initial microstate $0$ $\left[
p_{0}(0)=1,S(0)=0\right]  $ requires $p_{0}(t)=1$, for which all replicas must
be in the microstate $0$ simultaneously. This can occur in only one way.
However, such a true recurrence is impossible in stochastic systems. To show
this, we consider the situation in equilibrium; see (\ref{Eq_Probability}).
The number of possible ways the replicas can be arranged at time $t$,
consistent with microstate probabilities $p_{i}(t)$, is
\[
\mathcal{N}!/%
{\textstyle\prod}
\left[  \mathcal{N}p_{i}(t)\right]  !\approx e^{\mathcal{N}S(t)}%
=W^{\mathcal{N}},
\]
one of which is the true recurrent state. Hence, the probability for the
initial microstate to truly recur is $W^{-\mathcal{N}}\rightarrow0$ as
$\mathcal{N}\rightarrow\infty$. The recurrence of $0$ occurs
\cite{Lebowitz,Huang} several times, but with $p_{0}(t)<1$, so that other
microstates have to be considered to determine $S(t)$.

In conclusion, we have shown that the entropy of an isolated deterministic
system in a Poincar\'{e} cycle remains constant, so there is no violation of
the second law. Furthermore, we have also shown that the second law requires
that the probabilistic nature of microstates must be caused by external noise,
and not the mode of preparation alone. Once the entropy reaches its maximum
value, it remains constant. It never decreases. Thus, the second law is never violated.


\begin{thebibliography}{99}
\bibitem {Note}In classical physics, it is taken for granted as an axiom that
a precise measurement and preparation can always be made, at least in
principle, so that the state can be prepared in a unique microstate. In
quantum mechanics, the same is also true of an isolated system. A quantum
measurement, according to one of the axioms, will prepare the system in one of
the eigenstates; the latter are identified as the microstates.
\end{thebibliography}
\end{document}